\documentstyle[preprint,aps,prd]{revtex}
\begin{document}
\title{ Wess--Zumino terms and Duality}
\tightenlines

\author{A.Smailagic \footnote{E-mail address: a.smailagic@etfos.hr} }
\address{Department of Physics, 
Faculty of Electrical  Engineering \\
University of Osijek, Croatia }
\author{E. Spallucci\footnote{E-mail address:spallucci@trieste.infn.it}}
\address{Dipartimento di Fisica Teorica\break
Universit\`a di Trieste,\\
INFN, Sezione di Trieste}
\maketitle

\begin{abstract}
 We show the equivalence between St\"uckelberg and Wess--Zumino methods
  of restoration of gauge symmetries of the anomalous,
  Abelian, effective action, in arbitrary even dimensions $D=2k$. We present
   dual version of Wess--Zumino terms  with the  compensating field described 
   by a Kalb Ramond like  $p=2k-2$ form. 
  \end{abstract}
\newpage

By now we known that the anomalous behavior of classical symmetries
caused by  quantum matter effects is  encoded in the anomalous effective 
action either of gauge fields or gravity \cite{jack}, \cite{tilo}. 
Anomalous effective action reflects 
the incompatibility of various classical symmetries at the quantum level. 
Therefore, the best one can do quantum mechanically is to preserve one of the 
symmetry on the expense of the other in non-chiral models, while in chiral
models even this choice is not possible. 
  Various proposals to restore broken symmetries at the quantum level
have been put forward. One of these proposals, largely exploited in the 
literature, is based on {\it ad hoc} construction of additional {\it local} 
terms  in the anomalous effective Lagrangian, known as Wess--Zumino terms 
\cite{wz}, which restore the broken symmetry of the modified effective 
Lagrangian $L_{inv}=L_{eff}+L_{WZ}$.\\ 
On the other hand, another approach, apparently unrelated to Wess--Zumino idea,
was introduced in order to make classical, {\it massive,} Proca theory 
gauge invariant, and is known as the St\"uckelberg compensating 
formalism \cite{stuck}.\\
We conjecture that both above-mentioned methods 
could be {\it equivalent.} Our motivation is that both methods achieve the 
same goal, i.e. to restore a broken gauge invariance of an {\it Abelian}
vector theory.\\
In this letter we would like to clarify the conjectured equivalence 
between St\"uckelberg and Wess--Zumino methods. Let us start by recalling the 
Wess--Zumino idea of restoring gauge invariance in anomalous effective gauge 
theories. To maintain  pedagogical transparency we shall start from the simple
and well known $2$D effective anomalous Lagrangian of the Schwinger model 
\cite{jackiw} which is given by 

\begin{equation}
L_{eff}=-{ e^2\over 4\pi }\, \left[\, F_{\mu\nu}{1\over\Box }\,
 F^{\mu\nu}-2{\bf a}\, A^{\mu}A_{\mu}\,\right]\label{gamma1}
\end{equation}

(\ref{gamma1}) was obtained from a Dirac fermion  matter Lagrangian,  
having both vector and axial $U(1)$, {\it  classical,} gauge symmetries.
In order to treat the quantum breaking of classical symmetries on the same 
footing we have introduced an arbitrary, regularization dependent,
parameter ${\bf a}$ \cite{anais} in the effective Lagrangian (\ref{gamma1}). 
The effective Lagrangian $L_{eff}$ varies, under  
gauge transformation $ \delta_\Lambda A_{\mu}=\partial_\mu \Lambda$, as

\begin{equation}
\delta_\Lambda L_{eff} = -{ e^2\over \pi }{\bf a}\,\Lambda(x) \,
\partial_\mu A^{\mu}\ne 0\label{vargamma}
\end{equation}

Eq.(\ref{vargamma}) shows the gauge non-invariance of the $2$D effective 
Lagrangian, which is the origin of the gauge anomaly.  One could object that the 
gauge 
invariance can be restored  by a suitable choice of the arbitrary parameter 
${\bf a}$ as ${\bf a}=0$.  However, this choice would simply shift the anomaly 
from the vector to the axial current.  We want to show how {\it both } axial
 and vector symmetries can be restored simultaneously at the quantum level, 
 for any choice of ${\bf a}$. 
 Therefore, we shall keep the parameter ${\bf a}$ arbitrary throughout the 
 paper.\\ 
In order to restore gauge invariance in (\ref{gamma1}),  Wess and Zumino
have proposed to construct  a suitable, local, term whose gauge variation
cancel the gauge variation of the effective Lagrangian. A straightforward guess 
for the Wess--Zumino Lagrangian would be

\begin{equation}
L_{WZ}\equiv {\bf a}{ e^2\over \pi }\, \phi(x) \,
\partial_\mu A^{\mu} 
\end{equation}

where, one introduces the scalar field $\phi(x)$ transforming as 
$\delta_\Lambda\phi(x)= \Lambda(x) $ under gauge transformations. 
However, careful examination of such  Wess--Zumino Lagrangian shows that it is 
gauge non-invariant, not only due to the non-invariance of $\phi$, but also
due to the explicit dependence on the gauge field  $A$. Therefore, additional 
kinetic term for $\phi$ field is needed to compensate  non-invariance coming
from the gauge field itself, leading to the correct form of  the $2$D 
Wess--Zumino Lagrangian for the Schwinger model:

\begin{equation}
L\equiv -{ e^2\over 4\pi }\,\left[\, -4{\bf a}\, \phi(x) \,
\partial_\mu A^{\mu} + 2{\bf a}\,\phi\, \Box\, \phi \,\right]
\end{equation}

Once the modified effective Lagrangian $L_{inv}$ is defined as 

\begin{equation}
L_{inv}\equiv L_{eff} +  L_{WZ} =-{ e^2\over 4\pi }\, \left[\,
F_{\mu\nu}{1\over\Box }\, F^{\mu\nu}
-2\,{\bf a}\, A^{\mu}A_{\mu}\,   -4\,{\bf a}\,\phi(x) \,
\partial_\mu A^{\mu} + 2\,{\bf a}\,\phi\,\Box\,\phi \,  \right] 
\label{gammainv}
\end{equation}

then, one can verify that the addition of  Wess--Zumino piece ensures the gauge 
invariance of the total effective Lagrangian  for {\it any choice} of the 
parameter ${\bf a}$. In this way, the Wess--Zumino idea achieves its goal. 
On the other hand, a quick look at  (\ref{gammainv}) shows that it can be 
re-written as 

\begin{equation}
L_{inv}=-{ e^2\over 4\pi }\, \left[\, F_{\mu\nu}{1\over\Box }\,
 F^{\mu\nu} -2\,{\bf a}\,(\, A^{\mu}-\partial^\mu\phi\,)
(\, A_{\mu}-\partial_\mu\phi\,)\,\right]\label{gammainv2}
\end{equation}

Thus,  starting from gauge non-invariant Lagrangian 
(\ref{gamma1}), and making it gauge invariant following Wess--Zumino
prescription, we got the invariant effective Lagrangian (\ref{gammainv2}), 
which is nothing else but the gauge invariant massive Lagrangian originally 
proposed by St\"uckelberg for Proca theory. Reversing
the conclusion, we have shown that the advocated St\"uckelberg origin of the 
Wess--Zumino terms leads to the {\it equivalence} of the two methods.  \\
Encouraged by this toy model, one can proceed further and 
generalize the above result to {\it any} Abelian anomalous gauge theory. 
Notice that the relation between Wess--Zumino terms in the effective Lagrangian 
and St\"uckelberg mass term is specific to $2$D since it is only
in $2$D that a quantum anomaly induces a mass term for the gauge field. 
Anomalies in higher dimensions do not give origin to the mass term, but a 
simple generalization of the  St\"uckelberg idea allows to proceed along the 
same line as in $2$D to prove the equivalence of the Wess--Zumino and 
St\"uckelberg approach. In fact, St\"uckelberg  compensation of   
non-invariant,  Abelian, vector theories is achieved by the substitution of 
the complete field $A_\mu$ with the combination  $A_\mu - \partial_\mu \phi$, 
where $\phi$ is a compensating scalar field transforming as 
$\delta_\Lambda\phi=\Lambda$. from the previous discussion it follows that
to implement the compensation mechanism one needs the {\it explicit} form of the 
anomalous effective Lagrangian. Recently, we have obtained the general form of 
the anomalous Lagrangian  in arbitrary even dimensional ($D\equiv 2k$) 
space-time: 

\begin{eqnarray}
L_{eff}=\sum_{m=0}^{m_{max}}{ g^{2m+1} e^{k-2m}\over (2\pi)^k }\,  
\epsilon^{\mu_1\dots \mu_{2k-4m}\nu_1\dots\nu_{4m} }
F_{\mu_3\mu_4}\dots
F_{\mu_{2k-4m-1}\mu_{2k-4m}}&&\nonumber\\
\times F_{\nu_1\nu_2}^5\times\dots\times F_{\nu_{4m-1}\nu_{4m}}^5
\left[\, F_{\mu_1\mu_2}{1\over\Box }
\left(\,\partial^\mu A^{5}_\mu\,\right)  
- 2{\bf a}\,  A_{\mu_1}A^5_{\mu_2}\,\right]&&\label{effact}
\end{eqnarray}

where, $e$ and $g$ are the vector and axial coupling constants. Except than
in the very special $2D$ case, both the vector field $A_\mu$ and the axial 
field $A_\mu^5$ are independent gauge potentials, corresponding to anomalous 
local symmetries.  $m_{max}$ is a maximal integer number compatible with the 
restriction $0\le m \le (k-1)/2$, imposed by the anomalous Feynman diagrams in 
even dimensions. Details can be found in Ref.\cite{noilast}. As a check, one can
obtain the  Schwinger model anomalous effective Lagrangian (\ref{gamma1}) by 
putting $k=1$, $m=0$ in (\ref{effact}) and exploiting the  $2D$ relation 
$A_\mu=\epsilon_{\mu\nu}A^\nu{}^5$. \\
Now we are ready to address the St\"uckelberg derivation of  Wess--Zumino terms
in full generality. A gauge invariant St\"uckelberg Lagrangian can be obtained
from (\ref{effact}) by replacing $A_\mu$ and $A^5_\mu$ with the corresponding
compensated potentials: 

\begin{eqnarray}
&&L_{inv}^{Stuck.}=\sum_{m=0}^{m_{max}}{ g^{2m+1} e^{k-2m}\over (2\pi)^k }\,  
\epsilon^{\mu_1\dots \mu_{2k-4m}\nu_1\dots\nu_{4m} }
F_{\mu_3\mu_4}\dots
F_{\mu_{2k-4m-1}\mu_{2k-4m}}\nonumber\\
&&\times F_{\nu_1\nu_2}^5\times\dots\times F_{\nu_{4m-1}\nu_{4m}}^5
\left[\, 
F_{\mu_1\mu_2}{1\over\Box }
\partial^\mu\left(\, A^{5}_\mu- \partial_\mu\phi^5\,\right)\,
- 2\,{\bf a}\, 
\left(\,  A_{\mu_1}- \partial_{\mu_1}\phi\,\right)
\left(\, A^{5}_{\mu_2}  - \partial_{\mu_2}\phi^5\,\right)   
\,\right]\label{effstuck}
\end{eqnarray}

Dealing with two independent vector and axial vector gauge fields, one needs
{\it two, independent} St\"uckelberg compensators $\phi$ and $\phi^5$. 
Straightforward manipulations of the above Lagrangian 
lead to the equivalent Wess--Zumino form:

\begin{eqnarray}
L_{inv}&&= L_{eff} + L_{WZ} \label{linv}\nonumber\\
&&=L_{eff} +\sum_{m=0}^{m_{max}}{ g^{2m+1} e^{k-2m}\over (2\pi)^k }\,  
\epsilon^{\mu_1\dots \mu_{2k-4m}\nu_1\dots\nu_{4m} }
F_{\mu_3\mu_4}\dots
F_{\mu_{2k-4m-1}\mu_{2k-4m}}\nonumber\\
&\times & F_{\nu_1\nu_2}^5\times\dots\times F_{\nu_{4m-1}\nu_{4m}}^5
\left[\,
({\bf a}-1)F_{\mu_1\mu_2}\,\phi^5 -{\bf a}\,
F_{\mu_1\mu_2}^5\,\phi\,\right]\nonumber\\
&&\equiv L_{eff} - \left[\,
 X^{5\,\rho}\,\partial_\rho\, \phi^5 + X^\rho\,\partial_\rho\, \phi\,\right]
\label{nove}
\end{eqnarray}
where, we neglected surface terms.
In the Wess--Zumino part of the Lagrangian we have introduced  the  following
definitions \footnote{ Eq.(\ref{x}) and eq.(\ref{x5}) are defined as the most
 general expressions which allow to transform $\epsilon \phi F$
   into $X\partial\phi$ eq.(\ref{nove}).
  Dependence on the parameter $m$ counts the number of ways in which
  this can be done.}
 
 \begin{eqnarray}
 X^{5\,\rho} &&\equiv 2({\bf a} -1) 
\sum_{m=0}^{m_{max}}{ g^{2m+1} e^{k-2m}\over (2\pi)^k }
\epsilon^{\rho\mu_2\dots\mu_{2k-4m}\nu_1\dots\nu_{4m}}  
F_{\mu_3\mu_4}\dots
F_{\mu_{2k-4m-1}\mu_{2k-4m}}\nonumber\\
&&\times F_{\nu_3\nu_4}^5\dots F_{\nu_{4m-1} \nu_{4m}}^5\left[\,
A_{\mu_2}F_{\nu_1\nu_2}^5 + {2m\over k}\left(\, A_{\mu_2}^5 F_{\nu_1\nu_2}-
A_{\mu_2}F_{\nu_1\nu_2}^5\right)\right]
\,\label{x5}
\end{eqnarray}

and

\begin{eqnarray}
 X^\rho \equiv &- & 2{\bf a} 
\sum_{m=0}^{m_{max}}{ g^{2m+1} e^{k-2m}\over (2\pi)^k  }
 \epsilon^{\mu_1\dots\mu_{2k-4m}\nu_1\dots\nu_{4m}} 
F_{\mu_5\mu_6}\dots
F_{\mu_{2k-4m-1}\mu_{2k-4m}}F_{\nu_3\nu_4}^5\dots F_{\nu_{4m-1} \nu_{4m}}^5
\nonumber\\
&\times &\left[\,
\delta^\rho_{\mu_3}\, F^5_{\nu_1\nu_2}\, A_{\mu_4} F_{\mu_1\mu_2}^5 +{2m+1\over
k}\left(\,
\delta^\rho_{\mu_1}\, A_{\mu_2}^5\,  F_{\mu_3\mu_4}\, F^5_{\nu_1\nu_2} -
\delta^\rho_{\nu_1}\, A_{\nu_2}\,  F_{\mu_3\mu_4}^5 F_{\mu_1\mu_2}^5 \,\right)
\,\right]
\label{x}
\end{eqnarray}

One indeed recognizes  the  additional terms in (\ref{nove}) as Wess--Zumino 
terms. This gives the proof of equivalence of the two methods for arbitrary 
spacetime dimension. For the record, only the term $X^{5\,\mu}$ with 
 ${\bf a}=0$ can be found in the literature. One is indeed 
free to preserve at the quantum level one of the two classical symmetries by
a proper choice of the regularization method. Accordingly, 
the anomaly is conventionally shifted in the divergence of the axial current. 
However, we have proven that one  can restore {\it both} symmetries 
simultaneously, for any value of ${ \bf a}$, through the construction of 
independent Wess--Zumino terms. This result may be of particular importance  
for chiral models where { \it none} of the classical symmetries can be 
preserved at the quantum level \cite{jraja}.\\
 In a  recent paper we have studied a Proca theory where gauge symmetry
 is explicitly broken by a {\it classical,} mass term.  This theory has been 
 made gauge invariant by the introduction
 of the St\"uckelberg scalar compensator, and proven to be  dual to the  
 $B\wedge F$ model \cite{noibf}.  On the other hand, in the first part of the 
 present paper we have extended the notion of St\"uckelberg compensation to 
 gauge non--invariant anomalous theories at the quantum level, and proven the 
equivalence between the St\"uckelberg and Wess--Zumino methods. Combining
the  results  of  the present paper and the ones in\cite{noibf},  we conjecture
that the duality ideas of the latter should apply to the former. We mean
 that  the scalar compensator in the Wess--Zumino  terms,  
 in (\ref{nove}),  can be dualized to a Kalb--Ramond-like $p=2k-2$ forms. 
 In order to achieve this we
 adopt  general dualization procedure described in \cite{noibf}. One  
 starts from a suitably chosen  parent Lagrangian $L_P$ which, in the present 
 case, turns out to be of the  form \footnote{We assign canonical dimensions 
 (in units of mass) as 
follows: $[\, A\,]=k-1$,  $[\,\phi\, ]=k-2$.\\ This choice implies  dimensional
 coupling constants   $[\, e\, ]=[\, g\, ]=2-k $, and Wess--Zumino 
 terms $[\,\partial X\, ]=k+2$. Accordingly,  kinetic terms have
 dimension $2k$. Moreover, in the parent Lagrangian we require $[\,H\,]=k$.}

\begin{eqnarray}
L_P =&- &{1\over 2(2k-1)!} H _{\mu_1\dots\mu_{2k-1}}\, H^{\mu_1\dots\mu_{2k-1}}
+{M\over (2k-1)!} \epsilon^{\mu_1\dots\mu_{2k}} A_{\mu_{2k}} \, 
H_{\mu_1\dots\mu_{2k-1}}
\nonumber\\
&-&{ 1\over (2k-1)!}\left(\, M\, H^{\mu_1\dots\mu_{2k-1}} +{1\over f}
 X^{\,\mu_1\dots\mu_{2k-1}}\,
\right)\, F^*{}_{\mu_1\dots\mu_{2k-1} }(\,\phi\,)  \nonumber\\
&- &{1\over 2(2k-1)!} H^{5\, }_{\mu_1\dots\mu_{2k-1}}\, H^{5\,\mu_1\dots
\mu_{2k-1}}
+{M_5\over (2k-1)!} \epsilon^{\,\mu_1\dots\mu_{2k}}A_{\mu_{2k}}^5  \, 
H^{5\,}_{\mu_1\dots
\mu_{2k-1}}
\nonumber\\
&-&{1\over (2k-1)!}\left(\, M_5\, H^{5\,\,\mu_1\dots\mu_{2k-1}} +{1\over f_5}
X^{5\,\mu_1\dots\mu_{2k-1}}\,
\right)\, F^*{}_{\mu_1\dots\mu_{2k-1} }(\,\phi^5\,)
\label{parent}
\end{eqnarray}

where, $X^{\mu_1\dots\mu_{2k-1}}$ and $X^{5\,\mu_1\dots\mu_{2k-1}}$ are Hodge 
duals of (\ref{x}) and (\ref{x5}). $ \phi$ and  $\phi^5$  are 
St\"uckelberg scalar compensators which will be dualized to 
Kalb--Ramond-like fields. $H$ and $H^5$   are a priori {\it independent}
fields  in the parent Lagrangian (\ref{parent}).  We have also introduced 
dimensionless constants $f$, $f_5$ (numbers) in front of $X$ and $X^5$ in order 
to 
identify  {\it quantum } symmetry breaking terms at various stages of 
calculation. 
These factors are not present in the original Lagrangian (\ref{nove}). 
Thus, in order to achieve invariance of the anomalous theory we have to 
take $f=f_5=1$ when calculating appropriate variations. \\
It is worth mentioning the important point regarding the
construction of $L_P$. The parent Lagrangian must have certain symmetry
properties which will be reflected in the (dual) theories derived from it.
Let us require the gauge invariance of the complete Lagrangian $L_{inv}$ as
$\delta_\Lambda L_{inv}\equiv \delta_\Lambda L_P + \delta_\Lambda L_{eff}=0$
Then, it turns out that the variations of the terms involving classical mass
parameters cancel among themselves in (\ref{parent}). On the other hand, 
variations of the (quantum) terms involving dimensionless $f$ and $f_5$ cancel 
against the variations of the effective Lagrangian (\ref{effact}). \\
The dualization proceeds as  follows:
varying (\ref{parent}) with respect to $H$ we find the solution

\begin{eqnarray}
&& H^{\mu_1\dots\mu_{2k-1}}= M\,\left(\, \epsilon^{\mu_1\dots\mu_{2k-1}\mu_{2k}} 
A_{\mu_{2k}}
-F^*{}_{\mu_1\dots\mu_{2k-1} }(\,\phi\,) \,\right)\\
&& H^{5\,\, \mu_1\dots\mu_{2k-1}}= M_5\,\left(\, \epsilon^{\mu_1\dots\mu_{2k-1}
\mu_{2k}}
A_{\mu_{2k}}^5
-F^*{}_{\mu_1\dots\mu_{2k-1} }(\,\phi^5\,) \,\right)
\end{eqnarray}

and re-inserting the solution back into $L_P$ , we find the 
St\"uckelberg-like model 
  
\begin{eqnarray}
L_{V  A}= &-&{M^2\over 2}\, \left(\, A_\mu -\partial_\mu\phi\, \right)^2  
-{M_5^2\over 2}\, \left(\, A^{5\,}_\mu-\partial_\mu \phi^{5\,}\, \right)^2
\nonumber\\
&-&{1\over f\,(2k-1)!}
F^{\, *}_{\mu_1\dots\mu_{2k-1}}(\phi)\, X^{\mu_1\dots\mu_{2k-1}}
-{1\over f_5\, (2k-1)!}
F^{\, *}_{\mu_1\dots\mu_{2k-1}}(\phi^{5\,})\, X^{5\,\,\mu_1\dots\mu_{2k-1}}
\label{lstuck}
\end{eqnarray}
 
The first two terms in  the Lagrangian (\ref{lstuck}) are  gauge invariant mass
terms, while the last two terms correspond to Wess--Zumino terms of
(\ref{nove}). In this way both classical and quantum non-invariances have
been improved by the same scalar compensator $\phi$ transforming as
$\delta_\Lambda \phi = \Lambda$.\\
The Lagrangian dual to (\ref{lstuck})  is obtained by varying the
parent Lagrangian (\ref{parent}) with respect to both scalars $\phi$,
$\phi^5$. In this way we get

\begin{eqnarray}
&& H_{\mu_1\dots\mu_{2k-1}}= {1\over M}\,\left(\, \partial_{[\,
\mu_1}B_{\mu_2\dots\mu_{2k-1}\,]}- {1\over f} X_{\mu_1\dots\mu_{2k-1}}\,\right)
\label{hcl}\\
&& H^5_{\mu_1\dots\mu_{2k-1}}= {1\over M_5}\,\left(\, \partial_{[\,
\mu_1}B^5_{\mu_2\dots\mu_{2k-1}\,]}- {1\over f_5} X^5_{\mu_1\dots\mu_{2k-1}}\,
\right)\label{h5cl}
\end{eqnarray}

Re-inserting \ref{hcl} and \ref{h5cl} back into (\ref{parent}) we find the dual
Lagrangian:

\begin{eqnarray}
L_{dual}= &-&{1\over 2(2k-1)!\, M^2}\, \left(\, \partial_{[\, \mu_1}\,
B_{\mu_2\dots\mu_{2k-1}\,]}-{1\over f}X_{\mu_1\dots\mu_{2k-1}}\,\right)^2
\nonumber\\
&-&{1\over (2k-1)!}\,  \epsilon^{\mu_1\dots\mu_{2k}} \,
\left(\, \partial_{[\, \mu_1}\, B_{\mu_2\dots\mu_{2k-1}\,]}
 -{1\over f}\, X_{\mu_1\dots\mu_{2k-1}}\,\right)\,A_{\mu_{2k}}
\nonumber\\
&-&{1\over 2(2k-1)!\, M_5^2}\, \left(\,\partial_{[\, \mu_1}\,
B_{\mu_2\dots\mu_{2k-1}\,]}^5
-{1\over f_5}X^{5}_{\mu_1\dots\mu_{2k-1}}\,\right)^2 +\nonumber\\
&-&{1\over (2k-1)!}  \epsilon^{\mu_1\dots\mu_{2k}} \left(\,  
 \partial_{[\, \mu_1}\, B_{\mu_2\dots\mu_{2k-1}\,]}^5  -{1\over f_5}\,
X_{\mu_1\dots\mu_{2k-1}}^5\,\right)\, A^{5\,}_{\mu_{2k}} 
\label{lbd}
\end{eqnarray}

which can be suitably re-written as

\begin{equation}
L_{dual}=  L_{BF}+ L_{WZ}
\label{lduale}
\end{equation}

\begin{eqnarray}
L_{BF}= &&-{1\over 2(2k-1)!\, M^2}\,\left[\, H_{\mu_1\dots\mu_{2k-1} }(B)\,
\right]^2
-{1\over (2k-1)!}\,  \epsilon^{\mu_1\dots\mu_{2k}} \,H_{\mu_1\dots\mu_{2k-1}}
(B)\,A_{\mu_{2k}}\nonumber\\
&&-{1\over 2(2k-1)!\, M_5^2}\,\left[\, H_{\mu_1\dots\mu_{2k-1} }^5(B)\,\right]^2
-{1\over (2k-1)!}\,  \epsilon^{\mu_1\dots\mu_{2k}}\,
H_{\mu_1\dots\mu_{2k-1}}^5(B)\,A_{\mu_{2k}}^5\label{lbf}
\end{eqnarray}
\begin{eqnarray}
L_{WZ}=&&-{  \left(\, X_{\mu_1\dots\mu_{2k-1}}\,\right)^2\over 2 M^2f^2\,(2k-1)! 
} 
+ { X^{\mu_1\dots\mu_{2k-1}} \over f\,(2k-1)!}\,\left(\,
\epsilon_{\mu_1\dots\mu_{2k-1}\mu_{2k}}A^{\mu_{2k}} +{1\over M^2 }\,
 H_{\mu_1\dots\mu_{2k-1}}(B)\,\right) 
\nonumber\\
&&-{ \left(\, X_{\mu_1\dots\mu_{2k-1}}^5 \,\right)^2 \over 2M_5^2f_5^2\, (2k-1)! 
} 
+ { X^{5\,\,\mu_1\dots\mu_{2k-1}} \over f_5\,(2k-1)!}\,
\left(\,
 \epsilon_{\mu_1\dots\mu_{2k-1}\mu_{2k}}A^{5\,\,\mu_{2k}} + {1\over M_5^2 }\,
 H_{\mu_1\dots\mu_{2k-1}}^5(B)\,\right) \nonumber\\
 \label{lbd2}
\end{eqnarray}

 $L_{dual}$ has been split in two terms: $L_{BF}$ corresponds
 to the classical $B\wedge F$ (without quantum anomaly), which is dual to the
 mass term in (\ref{lstuck});  $L_{WZ}$ provides the dual version of the
 Wess--Zumino terms  in (\ref{lstuck}). 
Thus, (\ref{lbd}) and (\ref{lstuck}) are dual to each other, and 
we can summarize the effects of the dualization as follows:\\
\noindent
\underline{{\it Classical St\"uckelberg model}} 
Let us start without  Wess--Zumino terms, or quantum anomaly in the effective 
Lagrangian. Only classical  mass terms $M$ and $M_5$ break gauge symmetries. 
They are made invariant by the scalar  compensator  $\phi$ a la'  St\"uckelberg 
in
(\ref{lstuck}). Their dual form is given by (\ref{lbf}) in terms of the
  $B\wedge F$ model, in agreement with the  results in \cite{noibf}. \\
\noindent
\underline{{\it Quantum St\"uckelberg model}} 
Once classical mass terms are made gauge invariant, we switch on the quantum
anomaly. In (\ref{lstuck}) the quantum non-invariance of the effective
Lagrangian is improved by the Wess--Zumino terms expressed through
 $\phi$ and $\phi^5$. In the dual version
of the theory (\ref{lbd2}) the scalar compensators $\phi$ and  $\phi_5$ are 
replaced by the rank $p=2k-2$ Kalb--Ramond field $B$.
One can see from (\ref{lbd}) that the Kalb--Ramond-like field {\it is } in fact
 a St\"uckelberg compensator of $X(A)$ in the dual version of the theory. 
 As a compensator the $B$ field must vary under vector and axial 
 vector gauge transformation \footnote{
There is also an additional tensor gauge transformation of the $p=2k-2$
Kalb--Ramond fields described as 
\begin{eqnarray}
&&\delta B_ {\mu_1\dots\mu_{2k-2}} =
\partial_{[\,\mu_1}\Omega_{\mu_2\dots\mu_{2k-2}\,]}\nonumber\\
&&\delta B_ {\mu_1\dots\mu_{2k-2}}^5 =
\partial_{[\,\mu_1}\Omega_{\mu_2\dots\mu_{2k-2}\,]}^5\nonumber
\end{eqnarray}
which is still an invariance of the dual Lagrangian, and it is {\it anomaly 
free}.  Furthermore, in case of  classical duality
there is no need for additional gauge transformations of $B$ since
$B\wedge F$ term is automatically gauge invariant.}. 
On general grounds, if $\delta H(B)= f^{-1}\delta X$ and
$\delta X=\partial\left(\, \Lambda \,
 G(A, A^5)\,\right) $, then,  variation of Kalb--Ramond-like field is
  $ \delta B =f^{-1}\Lambda G(A, A^5)$. 
 The explicit form of $G(A, A^5)$ can be extracted from
 the transformations of  $B$ under vector and axial vector symmetry.
 From (\ref{x}) and (\ref{x5}) one finds

\begin{eqnarray}
\delta_\Lambda B_{\alpha_3\dots\mu_{2k}}= -{2{\bf a}\over f }\Lambda
\sum_{m=0}^{m_{max}}&&{ g^{2m+1} e^{k-2m}\over (2\pi)^k (2k-2)!} 
\delta^{[\, \mu_3\dots\mu_{2k-4m}\nu_1\dots\nu_{4m}\,]}
_{[\,\alpha_3\dots\alpha_{2k-4m} \dots\alpha_{2k}\,] }
\,\left(\, 1-{2m+1\over k}\,\right)
\times\nonumber\\ 
&&\times F_{\mu_3\mu_4}^5 F_{\mu_5\mu_6}\dots F_{\mu_{2k-4m-1}\mu_{2k-4m}}\,
  F^5_{\nu_1\nu_2}\dots F_{\nu_{4m-1} \nu_{4m}}^5\label{t1}
\end{eqnarray}

\begin{eqnarray}
\delta_{\Lambda_5} B_{\alpha_3\dots\mu_{2k}} = -{2{\bf a}\over f }\Lambda_5
\sum_{m=0}^{m_{max}}&&{ g^{2m+1} e^{k-2m}\over (2\pi)^k(2k-2)!} 
\delta^{[\,\mu_3\dots\mu_{2k-4m}\nu_1\dots\nu_{4m}\,]}_{[\,\alpha_3\dots
\alpha_{2k-4m}\dots\alpha_{2k}\,]}
\,\left(\, {2m+1\over k}
\,\right)
\nonumber\\ 
&&\times F_{\mu_3\mu_4} F_{\mu_5\mu_6}\dots F_{\mu_{2k-4m-1}\mu_{2k-4m}}\,
  F^5_{\nu_1\nu_2}\dots F_{\nu_{4m-1} \nu_{4m}}^5\label{t2}
\end{eqnarray}

\begin{eqnarray}
\delta_\Lambda  B^5_{\alpha_3\dots\mu_{2k}}   =  
{2({\bf a}-1)\over f_5}\,\Lambda
\sum_{m=0}^{m_{max}}&&{ g^{2m+1} e^{k-2m}\over (2\pi)^k(2k-2)!}
 \delta^{[\,\mu_3\dots\mu_{2k-4m}\nu_1\dots\nu_{4m}\,]}_{[\,\alpha_3\dots
 \alpha_{2k-4m}\dots\alpha_{2k}\,]}\,\left(\,1-{2m\over k}\,\right)
\nonumber\\
&&\times F_{\mu_3\mu_4} F_{\mu_5\mu_6}\dots F_{\mu_{2k-4m-1}\mu_{2k-4m}}\,
F^5_{\nu_1\nu_2}\dots F_{\nu_{4m-1} \nu_{4m}}^5\label{t3}
\end{eqnarray}

\begin{eqnarray}
\delta_{\Lambda_5}  B^5_{\alpha_3\dots\mu_{2k}}  =  {2({\bf a}-1)\over f_5}\,
\Lambda_5
\sum_{m=0}^{m_{max}}&&{ g^{2m+1} e^{k-2m}\over (2\pi)^k(2k-2)!}
 \delta^{[\,\mu_3\dots\mu_{2k-4m}\nu_1\dots\nu_{4m}\,]}_{[\,\alpha_3\dots
 \alpha_{2k-4m}\dots\alpha_{2k}\,]}\,\left(\,{2m\over k}\,\right)
\nonumber\\
&&\times F_{\mu_1\mu_2} F_{\mu_5\mu_6}\dots F_{\mu_{2k-4m-1}\mu_{2k-4m}}\,
F^5_{\nu_3\nu_4}\dots F_{\nu_{4m-1} \nu_{4m}}^5\label{t4}
\end{eqnarray}
 
 The above transformations guarantee the invariance of the complete Lagrangian
 $\delta L_{inv}\equiv \delta L_{eff}+\delta L_{dual}=0$, putting $f=f_5=1$. 
 Equations (\ref{lbd})
  and (\ref{lstuck}) show the dualization ``flipping''  $M\to 1/M$ and 
  $M_5\to 1/M_5 $, which forbids the massless  limit in the dual version of 
  the theory. In our model classical masses act as coupling constants in
 (\ref{parent}), and are reversed by the duality transformation. \\
For the sake of transparency, let us look at 
the four dimensional case ($k=2$, $m=0$) where (\ref{t1}), (\ref{t2}),
(\ref{t3}) and (\ref{t4}) give
 
  \begin{eqnarray}
&&\delta_\Lambda B_{\mu\nu}= -{{\bf a}\over f }\,
{ g e^2\Lambda\over (2\pi)^2} 
F_{\mu\nu}^5 
 \\
 &&\delta_{\Lambda_5} B_{\mu\nu}= {{\bf a}\over f }\,
{ g e^2\Lambda_5\over (2\pi)^2} 
\, F_{\mu\nu}  \\
 &&\delta_\Lambda B^5_{\mu\nu}= {{\bf a}-1\over f_5 }\,
{ g e^2\Lambda\over (2\pi)^2}\, 
\,F_{\mu\nu}
 \\
 &&\delta_{\Lambda_5} B^5_{\mu\nu}= 0\label{varb}
 \end{eqnarray}
 
 For the choice ${\bf a}=0$ the above formulae reproduce results of \cite{jap}. 
 One can verify that the gauge transformations of the
 Kalb--Ramond compensator in (\ref{varb}) agree with 
 the ones found in supergravity \cite{sugra}. This result
 suggests both a natural  SUSY extension of the gauge theory discussed
 in this paper, and gives further support to the role of the Kalb--Ramond 
 fields as a St\"uckelberg compensator, i.e. field transforming under vector and
  axial vector  gauge transformations.\\ 
 In summary, we have shown the {\it equivalence} of the 
 St\"uckelberg and Wess--Zumino methods  to restore the
  gauge invariance of  an {\it anomalous} Abelian theory of massive one-forms.
  Combining this equivalence with the results about classical duality 
  between St\"uckelberg and $B\wedge F$ theories,  we have found a new, dual, 
 form of Wess--Zumino terms. Massless 
  limits $M\to 0$ , $M_5\to 0$ cannot be performed as a consequence
  of the flipping between ``strong/weak coupling'' regimes. Thus, in order
  to produce dual theory one has to have classical mass terms form the start.
   \\    
  The application of the method described in this letter to non--Abelian, 
  anomalous gauge theories is currently under investigation. We are 
  developing a non--Abelian dualization procedure acting on Yang--Mills
  fields coupled to Kalb--Ramond tensors\cite{last}. What we are looking for 
  is the generalization of the  anomalous effective action (\ref{effact}) in the 
  non--Abelian case. Once this goal will be brought to a success, we shall 
  be able to build up the non--Abelian version of the model discussed in this
  letter.

\end{document}